\documentclass{iau}
\usepackage{graphicx,natbib}
 
\title{Understanding the transformation of spirals to lenticulars}

\author[Johnston, Arag\'on-Salamanca \& Merrifield]{Evelyn J.~Johnston$^1$, Alfonso~Arag\'on-Salamanca$^1$ and Michael~R.~Merrif\mbox{}ield$^1$}

\affiliation{$^1$School of Physics and Astronomy, University of Nottingham, University Park, Nottingham, NG7 2RD, UK\\
email: {\tt ejohnsto@eso.org} \\
}

\pubyear{2014}
\volume{309}
\jname{Galaxies in 3D across the Universe}
\editors{B. L. Ziegler, F. Combes, H. Dannerbauer, M. Verdugo, eds.}

\begin{document}

\maketitle

\begin{abstract}
By studying the individual star-formation histories of the bulges and discs of 
lenticular (S0) galaxies, it is possible to build up a sequence of events that leads 
to the cessation of star formation and the consequent transformation from 
the progenitor spiral. In order to separate the bulge and disc stellar populations, 
we spectroscopically decomposed long-slit spectra of Virgo Cluster S0s into 
bulge and disc components. Analysis of the decomposed spectra shows that
the most recent star formation activity in these galaxies occurred within the bulge 
regions, having been fuelled by residual gas from the disc. These results point 
towards a scenario where the star formation in the discs of spiral galaxies 
are quenched, followed by a final episode of star formation in the central regions
from the gas that has been funnelled inwards through the disc.

\end{abstract}
\firstsection
\section{Introduction}
Lenticular (S0) galaxies are often seen as an endpoint in the evolution of 
spiral galaxies since they display the same discy morphologies as spirals, 
but contain older stellar populations. 
%
In order to transform a spiral into an S0, the star-formation in the 
disc must be truncated and the bulge luminosity enhanced \citep{Christlein_2004}. 
Therefore, to understand how these two phenomena occur, we need to 
study the individual star-formation histories of the bulges and discs.

\section{Disentangling the Star Formation Histories of the Bulge and Disc}
A sample of 21 S0s from the Virgo Cluster were used in this study, with B-band magnitudes in the range of $-17.3$ to $-22.3$ and inclinations above 40 degrees to reduce contamination from misclassified ellipticals. A long-slit spectrum was obtained along the major axis of each galaxy using Gemini-GMOS with typical exposure times of \mbox{$\sim2-3$~hours}. The reduced spectra were decomposed using the spectroscopic bulge--disc decomposition technique of \citet{Johnston_2012}, in which light profile of the galaxy in each wavelength bin was separated into bulge and disc components by fitting a S\'ersic bulge plus exponential disc model. Having found the best fit and obtained the decomposition parameters for each wavelength bin, the total light from each component was then calculated by integration, and plotted against wavelength to create two one-dimensional spectra representing purely the bulge and disc light.

\begin{figure}
\centering
\includegraphics[width=0.9\columnwidth]{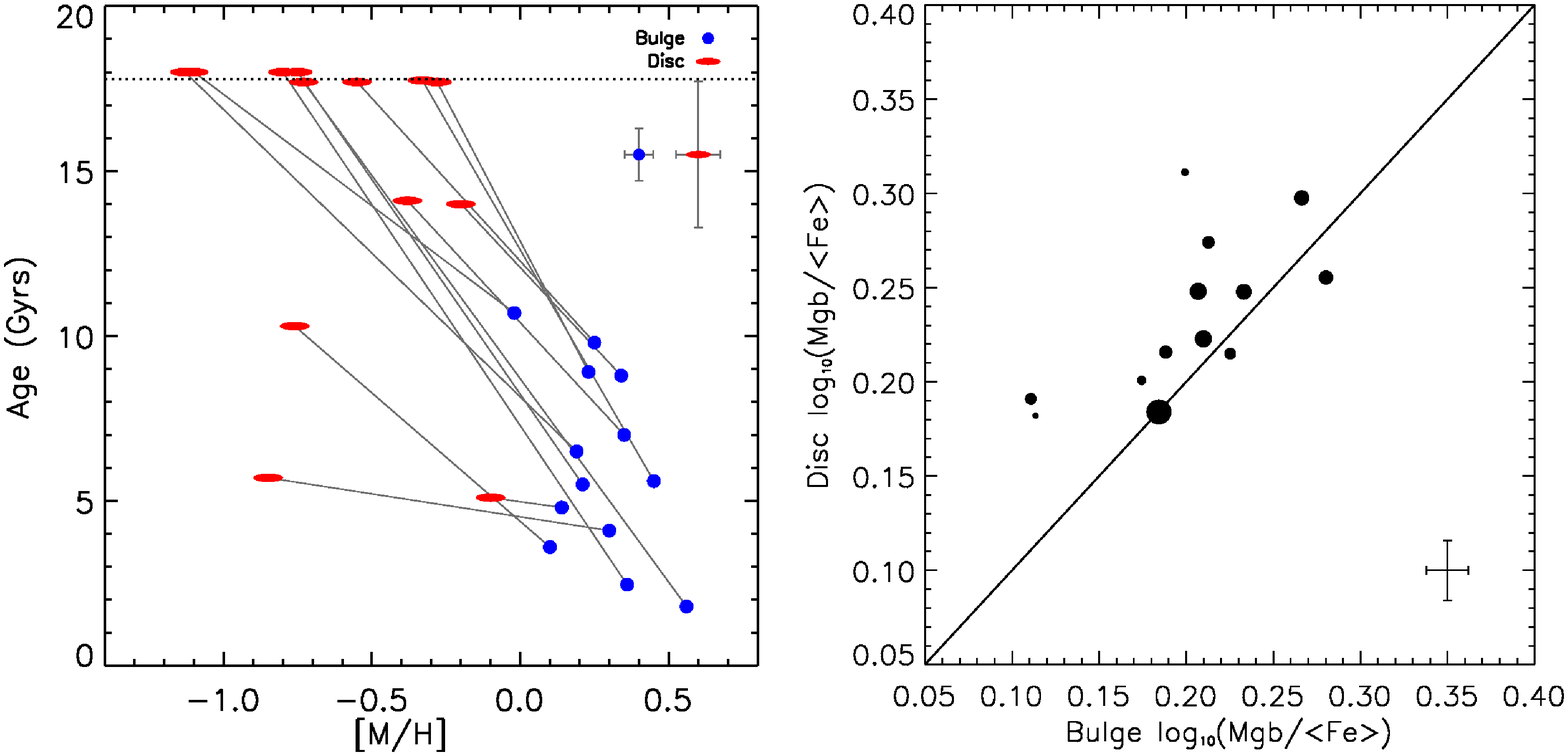} 
\caption{
\textit{Left}: The relative, light-weighted ages and metallicities of the bulges 
and discs of S0s in the Virgo Cluster. The lines link bulges and discs from the 
same galaxy.
\textit{Right}: Comparison of the Mgb/$\langle$Fe$\rangle$ ratios for the bulges 
and discs. Errors are shown on the right of each plot.
}\label{fig:Johnston_fig1}
\end{figure}

The ages, metallicities and Mgb/$\langle$Fe$\rangle$ ratios 
of the bulges and discs were measured using the strengths of the absorption features 
in the decomposed spectra. It can be seen in Fig.~\ref{fig:Johnston_fig1} that the 
bulges contain systematically younger and more metal rich stellar populations than 
their associated discs, suggesting that a final star-formation event occurred in the 
bulge region during the transformation. Additionally, a comparison of the 
Mgb/$\langle$Fe$\rangle$ ratios for the bulges and discs, also shown in 
Fig.~\ref{fig:Johnston_fig1}, reveals a correlation, indicating that their star-formation 
histories are linked. 

\section{Conclusions}

We find that during the transformation from a spiral to an S0,  gas is stripped 
from the disc gently enough that no significant amounts of star formation were 
triggered there. At some point, a final star-formation 
event occurs in the bulge region, resulting in younger, more metal rich stellar 
populations there. The most likely origin of the gas that fuelled this star formation 
is from the associated disc, which appears to be true from the clear 
correlation in the chemical enrichment of the the bulges and discs.
%
Together, these results present a scenario for the transformation of spirals to S0s in which 
gas is stripped gently from the disc, while the residual gas is channelled in towards 
the centre of the galaxy. Eventually, this gas induces a final episode of star formation 
within the bulge, which uses up all the remaining gas and boosts the luminosity of the 
bulge relative to the disc. The whole galaxy then fades to an S0. 

Clearly, further information on the transformation can be obtained if the 
decomposition was carried out in two
dimensions using integral field unit (IFU) data. The up-coming IFU Mapping of Nearby Galaxies 
at APO (MaNGA, Bundy et~al, Submitted to ApJ) survey in SDSS-IV
promises the resolution and wide field of view that will answer the remaining questions
about how the bulge and disc star-formation histories are linked.

\section*{Acknowledgements}

\noindent
EJ acknowledges support from the STFC, the IAU and the RAS.

\end{document}